\documentclass[11pt]{article}
\newcommand{\be}{\begin{equation}}
\newcommand{\ee}{\end
{equation}}
\newcommand{\bea}{\begin{eqnarray}}
\newcommand{\eea}{\end{eqnarray}}

\def\({\left(} \def\){\right)}
\usepackage{amssymb}
\usepackage{epsfig}
\usepackage{graphicx}
\begin{document}
\title {A Simplified Mathematical Model for the Formation of Null Singularities Inside
 Black Holes I -- Basic Formulation and a Conjecture} \maketitle
\author{  Amos Ori\footnotemark[1], Dan Gorbonos\footnotemark[2]\\

\footnotemark[1] Department of Physics\\
  Technion-Israel Institute of Technology\\
  Haifa 32000\\
  Israel \\
  amos@physics.technion.ac.il \\

\footnotemark[2]
 Racah Institute of Physics \\
 The Hebrew University \\
 Jerusalem 91904 \\
 Israel\\
{\tt gdan@phys.huji.ac.il}
\\

\begin{abstract}
Einstein's equations are known to lead to the formation of black
holes and spacetime singularities. This appears to be a
manifestation of the mathematical phenomenon of finite-time blowup:
a formation of singularities from regular initial data. We present a
simple hyperbolic system of two semi-linear equations inspired by
the Einstein equations. We explore a class of solutions to this
system which are analogous to static black-hole models. These
solutions exhibit a black-hole structure with a finite-time blowup
on a characteristic line mimicking the null inner horizon of
spinning or charged black holes. We conjecture that this behavior
--- namely black-hole formation with blow-up on a characteristic
line --- is a generic feature of our semi-linear system. Our simple
system may provide insight into the formation of null singularities
inside spinning or charged black holes in the full system of
Einstein equations.
\end{abstract}

\tableofcontents

\section{Introduction}

This paper examines a simple system of two equations inspired by the
Einstein equations. The main purpose is to gain insight into the
onset of null singularities inside spinning or charged black holes
(BHs).

To understand the background and motivation for our toy model it
will be worthwhile to review the development of our present
conception of the null singularity inside BHS. The $r=0$ curvature
singularity of the Schwarzschild geometry has been regarded for many
years as a prototype for the spacetime singularity expected to be
present inside BHs. However, the Reissner-Nordstrom (RN) solution,
describing a spherically symmetric charged BH, lacks a spacelike
$r=0$ singularity. Instead it admits an inner horizon (IH)--- a
perfectly smooth null hypersurface which constitutes a Cauchy
horizon (CH) for partial Cauchy surfaces outside the BH. The
(analytically extended) RN solution admits an $r=0$ singularity too,
but this singularity is timelike rather than spacelike, and it is
located beyond the IH (hence outside the Cauchy development). A
similar situation is found in the Kerr solution, describing a
stationary spinning BH: A perfectly smooth IH, which again functions
as a CH; and the spacetime singularity is timelike, located beyond
this null hypersurface. In both the RN and Kerr solutions, the
regular IH is known to be unstable to small perturbations, and this
instability leads to the formation of a curvature singularity
instead of a smooth IH. Thus, in order to explore the structure of
the singularities inside realistic spinning BHs, one must understand
the process of singularity formation due to the instability of the
IH.

Of the three BH solutions mentioned above---Schwarzschild, RN, and
Kerr---the one which is mostly relevant to realistic spinning BHs is
obviously the Kerr solution. Nevertheless there is a remarkable
similarity between the internal structures of spinning and charged
BHs, which allows one to use spherical charged BHs as a useful toy
model for the more realistic (but much more complicated) spinning
BHs.

The IH of the RN solution is the locus of infinite blue-shift, as
was already pointed out by Penrose~\cite{penrose}. Infalling
perturbations of various kinds are infinitely blue-shifted there,
which leads to instability of the IH~\cite{penrose1}. As a
consequence the latter becomes the locus of a curvature singularity,
to which we shall often refer as the {\it IH singularity}. In order
to explore this phenomenon, Hiscock~\cite{Hiscock} modeled the
blue-shifted perturbations by a null fluid---a stream of massless
particles. He analyzed the geometry inside a charged BH perturbed by
a single such stream, an ingoing null fluid, using the charged
Vaidya solution~\cite{vaidya}. He found that the IH becomes a
non-scalar null curvature singularity. Later Poisson and
Israel~\cite{poisson} explored the system of a charged BH perturbed
by two fluxes, namely both ingoing and outgoing null fluids. They
concluded that in this case too the IH becomes a null curvature
singularity. This time, however, the singularity is a
scalar-curvature one because the mass-function
---a scalar quadratic in derivatives of the area coordinate---diverges,
a phenomenon known as {\it mass-inflation}. The detailed structure
of this mass-inflation singularity was later analyzed \cite{Ori1}
within a simplified model (in which the outgoing flux is replaced by
a discrete null shell). This study showed that the metric tensor
(when expressed in appropriate coordinates) has a continuous and
non-singular limit at the singularity. Yet derivatives of the metric
functions diverge at the IH, yielding a curvature singularity. The
continuity of the metric has crucial physical consequences: It
implies that the singularity is weak~\cite{tipler}, namely an
extended object will only experience a finite (and possibly very
small) tidal deformation on approaching the IH singularity.

Subsequently more detailed numerical and analytical studies of the
mass inflation phenomenon were performed, in which the perturbations
were modeled by null fluids or by a self-gravitating scalar
field~\cite{Bonanno,Brady1,Burko1,Burko2,Burko3}. These studies
confirmed the conclusions of the earlier analyses
(\cite{Hiscock},~\cite{Ori1}). In addition, numerical analyses
revealed that, at least in the case of scalar field perturbations, a
spacelike singularity forms in the asymptotically-late advanced
time. More recently Dafermos~\cite{dafermos} proved for a
 characteristic initial value problem for the spherically symmetric
Einstein-Maxwell-Scalar Field equations that for an open set of
initial data on the event horizon (EH), the future boundary of the
maximal domain of development becomes a null surface along which the
curvature blows up. Dafermos proved that the metric can be
continuously extended beyond the IH, namely, the singularity is
weak.

The situation inside a spinning BH is similar in many aspects to
that of a spherical charged BH. Here, again, the inner horizon is
the locus of unbounded blue shift, suggesting that the regular IH of
the Kerr geometry will become a curvature singularity when
perturbed. A thorough perturbation analysis~\cite{Ori92}
\cite{OriLonger} showed that indeed a scalar-curvature singularity
forms at the early portion of the IH, which is again null and weak.
This picture of the spinning IH singularity was later confirmed by
an independent perturbative analysis by Brady et al.~\cite{Brady5}.
The existence of a class of solutions to the vacuum Einstein
equations which admit null, weak, scalar-curvature singularities was
also demonstrated in exact non-perturbative
analyses~\cite{planar,flanagan} (though these exact analyses, unlike
the earlier perturbative analyses,
 did not demonstrate the actual occurrence of a null weak singularity inside BHs).

If a cosmological constant $\Lambda>0$ is present, the spacetime is
no longer asymptotically flat and a cosmological horizon replaces
the future null infinity. The spherical charged BH and the
stationary spinning BH are then described by the
Reissner-Nordstr\"{o}m-de Sitter (RNDS) and Kerr-de Sitter
solutions, respectively. In both cases there are three horizons,
namely cosmological, event, and inner horizons. The surface gravity
of these horizons depend on the parameters of the solutions, namely
the cosmological constant, mass, and the charge or angular momentum.
We denote these surface gravities by $\kappa_{co}$ (cosmological
horizon), $\kappa_{ev}$ (EH) and $\kappa_{in}$ (IH). If
$\kappa_{in}>\kappa_{co}$, there is an infinite blue-shift at the
IH, suggesting an instability of the latter. This instability was
first investigated by Mellor and Moss~\cite{Mellor1} in the case of
spherical charged BHs, and by Chambers and moss~\cite{kerrds} for
spinning BHs, using linear perturbations in both cases. In the case
of a spherical charged BH the non-linear instability with respect to
ingoing null fluid was investigated by Brady and
Poisson~\cite{Brady2}. Brady, N\'{u}\~{n}ez, and Sinha~\cite{Brady3}
investigated a model in which both ingoing and outgoing null fluids
are present, and found that the mass function diverges provided that
$\kappa_{in}>2 \kappa_{co}$. In the range
 $2\kappa_{co}>\kappa_{in}>\kappa_{co}$ the mass function is finite,
yet the Kretchman curvature scalar
$R_{\alpha\beta\gamma\delta}R^{\alpha\beta\gamma\delta}$ diverges at
the inner horizon. Later, Brady, Moss, and Myers~\cite{Brady4}
considered also the contribution of the radiation that is scattered
by the curvature in the vicinity of the EH. They found that when
this scattering is taken into account, the necessary condition for
stability of the IH (namely bounded curvature) is that both
$\kappa_{co}$ and  $\kappa_{ev}$ are greater than $\kappa_{in}$.
None of the stationary (asymptotically-de Sitter) electro-vacuum
black holes satisfy this condition. Chambers~\cite{chambers} studied
a simplified mass-inflation model with a continuous ingoing null
fluid and a discrete outgoing shell and confirmed the earlier
results ~\cite{Brady3}. He also found that for all values of
$\kappa_{in}$ the metric functions are continuous and non-singular
at the IH, even though the mass function diverges. Namely, the
mass-inflation singularity is weak in the $\Lambda>0$ case as well.

The combination of all the above-mentioned investigations strongly
suggests (though a mathematical proof is still lacking) that the
vacuum (or electro-vacuum) Einstein equations admit a generic class
of solutions in which a null weak singularity forms inside a
spinning (or charged) black hole. In what follows we shall {\it
assume} that this is indeed the case. Now, the Einstein equation in
3+1 dimensions (and with the lack of any symmetry) is a rather
complicated non-linear dynamical system. The following question
therefore naturally arises: Is it possible to extract from the
Einstein equations a smaller and simpler dynamical system, which is
capable of producing black hole-like configurations with generic
null weak singularities inside them? If such a simpler system is
found, perhaps it could be viewed as the ``active ingredient'' of
the Einstein equations (as far as the formation of black holes and
null singularities is concerned). This may provide insight into the
mathematical process of the formation of null singularities. The
construction and exploration of such a simple system of equations is
our main goal in this paper.

The system of Einstein equations combines both evolution and
constraint equations. It appears likely, though, that the property
of producing generic null singularities is admitted by the
sub-system of evolution equations. We shall therefore extract our
simplified toy-system from the evolution equations and simply ignore
the constraint equations. Now, when the constraint equations are
discarded, one obtains dynamical behavior even in
spherically-symmetric situations (it is the constraint section which
``freezes'' the dynamics in spherical symmetry). Consequently, we
shall extract our toy-system from the evolution section of the
electro-vacuum Einstein equations in spherical symmetry.

We shall thus proceed in Sec. 2 as follows: We start from the
electro-vacuum Einstein equations in spherical symmetry. We also add
a cosmological constant, for reasons explained below. Then we
discard the constraint equations, and re-formulate the evolution
equations in a simple form free of first-order derivatives. This
yields a semi-linear hyperbolic system of two equations for the two
unknowns which we denote $R(u,v)$ and $S(u,v)$, where $u,v$ are two
null coordinate and $R,S$ are constructed from the metric functions
(specifically $R$ is the square of the area coordinate, and $s$ will
be specified below). The new system involves a ``generating
function'' $h(R)$ which in the above construction emerges in a very
specific form [see Eq. (\ref{speh})]. However, from the mathematical
point of view it appears likely that the global properties of the
solutions such as BH and singularity formation will not be sensitive
to the detailed functional form of $h(R)$, but only to certain
global and/or asymptotic features of this function. For this reason
we extend our view point and explore this semi-linear system with a
rather general function $h(R)$. This generalizes our investigation
and simplifies it at the same time.

Our strategy of considering a general function $h(R)$ also has a
side benefit: As it turns out, certain two-dimensional
general-relativistic dilatonic models can be re-formulated such that
their evolution sector is described by our semi-linear system, with
a certain function $h(R)$. This includes the model by Callan et
al.~\cite{CGHS} and its charged
generalization~\cite{Ori2,thorlacius2,thorlacius}. We describe this
at the end of Sec. 2.

In Sec. 3 we describe some basic mathematical properties of our
semi-linear system, including conserved fluxes, a generalized mass
function, and gauge freedom. The latter means that the semi-linear
system is invariant under coordinate transformations of the form
$u\to u'(u),v\to v'(v)$. Then in Sec. 4 we construct,  for any
$h(R)$, a class of exact solutions with vanishing fluxes, to which
we shall refer as the {\it flux-free solutions}. This is a
one-parameter family of solutions (for given $h(R)$), a
generalization of the RNDS solution to arbitrary $h(R)$. We then
observe that for functions $h(R)$ admitting three roots (or more),
the corresponding flux-free solution describes a RNDS-like styatic
BH, with three horizons, namely three null lines of constant $R$: an
event horizon located at a line $u=const$, and cosmological and
inner horizons, both located at $v= const$. (In the
``Eddington-like'' coordinates, in which the flux-free solution is
first derived, the three horizons are located at infinite value of
the relevant null coordinate, but this is later fixed by a
coordinate transformation as described below.) The three horizons
intersect at a single point P, representing the timelike infinity
for the external region between the cosmological and event horizons.
The function $s$ diverges at P, but this divergence does not
represent a spacetime singularity: Instead it reflects the fact that
the
 proper-time interval between P and any point to its past is infinite.
 The function $R$ is many-valued at P.

The singularity structure of the flux-free solution is studied in
Sec. 5. The solution becomes singular at the horizons ($s$
diverges), but this is merely a coordinate singularity. To
regularize the solution we transform $u,v$ to new, ``Kruskal-like'',
coordinates. (Specifically we ``Kruskalize'' $u$ with respect to the
event horizon and $v$ with respect to the cosmological horizon, such
that the initial data for the BH formation are regular.) In these
new coordinates the solution extends smoothly into the BH, and
provides a description of the internal geometry up to the inner
horizon.

The asymptotic form of the functions $R,s$ near the IH is the
primary objective of this paper. In the flux-free solution
(expressed in Kruskal coordinates and extended into the BH as
described above), $R$ admits a constant finite value along the IH
but $s$ diverges there (for a generic $h(R)$). This divergence, too,
does \textit{not} indicate a true singularity, because it can be
removed by ``Kruskalizing'' $v$ with respect to the inner horizon.
With such a coordinate transformation, the variables $R,s$ become
perfectly regular (in fact analytic) in the IH neighborhood.
\footnote{Note, however, that such a new ``Kruskalization'' will
spoil the original ``Kruskalization'' of $v$ at the cosmological
horizon, which will be expressed by a divergence of $s$ along the
latter.} In fact, this divergence of $s$ reflects the infinite
blue-shift (or red-shift in some cases) which takes place at the IH,
just as in the standard RN and RNDS geometries. It should be noted
that all invariant quantities involving the variables $R,s$ and
their derivatives are regular at the IH in the flux-free solution.

Consider now the initial-value problem for our semi-linear system.
The initial hypersurface is taken to be a spacelike hypersurface
which intersects both the event and the cosmological horizons. In
the first stage we assume that the initial data agree with those of
the flux-free solution. Then the evolving solution will be just the
flux-free solution, with an IH of the form described above. We
assume that the initial data are everywhere regular, which means
that the flux-free solution is obtained not in the Eddington-like
coordinates, but in other coordinates which are regular at the event
and cosmological horizons---e.g. the above mentioned Kruskal-like
coordinates.

The major challenge is now to understand how will the functions
$R,s$ be affected if the initial data are modified such that they no
longer agree with those corresponding to the flux-free solution.
What features of the BH and the IH will survive the perturbation,
and which features will be modified? We do not have a full answer to
this question, but we do have a conjecture that we present in Sec.
6, based on several compelling indications. These include the linear
perturbation of the flux-free solution, the ``generalized Vaidya
solution'' (valid for arbitrary $h(R)$; see the appendix), and also
some specific examples of $h(R)$ for which the general solution may
be constructed. Our conjecture may be stated very briefly as
follows: First, the global black-hole structure is unchanged (this
is manifested by the persistent divergence of $s$ at a point P where
the three horizons meet); Second, $R$ remains finite (though no
longer constant) along the IH; Third, the divergence of $s$ on
approaching the IH persists and preserves its leading asymptotic
form; and after ``re-Kruskalization'' $s$ becomes finite along the
IH, just as in the unperturbed flux-free solution. However, one
important difference occurs due to the deviation from the flux-free
initial data: Although the variables $R,s$ are continuous (after
``re-Kruskalization'') at the IH, they are no longer smooth. Certain
invariant quantities involving the derivatives of $R$ now diverge on
the IH (this holds provided that the ``surface gravity'' of the IH
is sufficienly large).

The $h(R)$ function corresponding to the electro-vacuum solutions in
four dimensions without a cosmological constant, namely Eq.
($\ref{speh}$) with $\Lambda =0$, has only two roots. The
corresponding BH solution has two horizons, the event and inner
horizons. The cosmological horizon disappears when $\Lambda $
vanishes, and instead there is a future null infinity. Our
semi-linear system is useful in this case too, but the initial-value
problem described by this system is conceptually more complicated in
this case. To understand the reason, consider a black-hole solution
with a cosmological horizon, and consider an initial spacelike
hypersurface $\Sigma$ which intersects both the event and
cosmological horizons. We can pick a compact portion $\Sigma_{0}$ of
$\Sigma$ which still intersects the event and cosmological horizons.
Then the early portion of the IH is included in the closure of
$D_{+}(\Sigma_{0})$, where $D_{+}$ denotes the future domain of
dependence. On the other hand, in the analogous asymptotically-flat
case the initial hypersurface $\Sigma$ must extend to spacelike
infinity (or alternatively to future null infinity) in order to have
any portion of the IH being included in the closure of
$D_{+}(\Sigma)$. This means that the behavior of $R,s$ near the IH
will depend on the asymptotic behavior of the initial data as the
initial hypersurface approaches spacelike infinity. No such
complication occurs in the case of a BH with a cosmological horizon:
Here it is sufficient to require that the initial data are
sufficiently regular on $\Sigma_{0}$, and the issue of their
large-$R$ asymptotic behavior does not arise. For this reason, we
shall restrict our attention in this paper to functions $h(R)$ with
at least three roots. Note that in this case the divergence of $s$
at P and at the IH is, from the PDE point of view, a manifestation
of the \textit{finite-time blow-up} phenomenon, caused by the
non-linearity of the hyperbolic system. (The standard
General-Relativistic point of view is somewhat different, however,
because the proper-time distance of P is infinite, due to the
divergence of $s$, so this divergence is not a spacetime
singularity.)

As was mentioned above, we view our semi-linear system as a toy
model for the much more complicated system of Einstein equations in
four dimensions. Obviously not all properties of the Einstein
equations are mimicked by our semi-linear system. The properties we
expect our toy system to display are (i) the very formation of the
BH (expressed in our system by the finite-time blow-up of $s$ at the
point P), (ii) the no-hair properties of the BH---namely the decay
of external perturbations, and (iii) the generic formation of a
null, scalar-curvature, weak singularity on the IH. We do not expect
our toy system to properly address the spacelike singularity (which
may intersect the IH at later times). Also this simple system is
incapable of describing the oscillatory character of the null IH
singularity inside a generically-perturbed spinning BH~\cite{Ori99}.
Nevertheless our toy system correctly demonstrates the basic
properties of the IH singularity---in particular its weakness.

We conclude in Sec. 7 by outlining some directions for future
research.

\section{The Field equations}
\label{field}

\subsection{Maxwell-Einstein equations in spherical symmetry}

We start by considering the Maxwell-Einstein equations in a
spherically symmetric spacetime. We write the metric in double-null
coordinates as
\begin{equation}
ds^{2}=-2f(u,v)dudv+r^{2}(u,v)d\Omega^{2}, \label{metric1}
\end{equation}
where $d\Omega^{2}=d\theta^{2}+\sin^{2}\theta d\phi^{2}$. The
Maxwell equations are easily solved, yielding
\[
F_{,uv}=-F_{,vu}=Qf/r^{2}
\]
with all other components vanishing. Here $Q$ is a free parameter,
to be interpreted as the charge. The electromagnetic energy-momentum
tensor
\[
T_{\mu\nu}=\frac{1}{4\pi}\left(g^{\alpha\beta}
F_{\mu\alpha}F_{\nu\beta}-\frac{1}{4}g_{\mu\nu}
F_{\alpha\beta}F^{\alpha\beta}\right)
\]
is then substituted in the Einstein equations with a cosmological
constant $\Lambda$,
\begin{equation}
G_{\mu\nu}+\Lambda g_{\mu\nu}=8\pi T_{\mu\nu}. \label{Einstein}
\end{equation}
This yields a system of two evolution equations, \be
r_{,uv}=-\frac{r_{,u}r_{,v}}{r}-\frac{f}{2r}\left(
1-\frac{Q^{2}}{r^{2}} -\Lambda r^{2}\right) \ee \be
f_{,uv}=\frac{f_{,u}f_{,v}}{f}+2\frac{f}{r^{2}}r_{,u}r_{,v}+\frac{f^{2}}{
r^{2}}\left( 1-2\frac{Q^{2}}{r^{2}} \right) \ee and two constraint
equations, \footnote{This involves a slight abuse of the standard
terminology as the notion of constraint equations is usually
formulated with respect to foliations of spacetime by spacelike
hypersurfaces. The terminology we use here is a natural extension of
the standard one to the double-null set-up in two effective
dimensions.} \be r_{,uu}=r_{,u}f_{,u}/f\;,\quad
r_{,vv}=r_{,v}f_{,v}/f\, \label{constrf}. \ee The latter two
equations (unlike the two evolution equations) are in fact ordinary
equations along the lines $v=const$ or $u=const$, respectively.

\subsection{Constructing our semi-linear system}
\label{semi} We first re-formulate the field equations such that no
first-order derivatives appear in the evolution equations. To this
end we introduce two new variables $R$ and $s$ instead of $r$ and
$f$:
\begin{eqnarray}
R\equiv r^{2}, & e^{s}\equiv rf \label{vara}.
\end{eqnarray}
With the new variables, the evolution equations take the convenient
form \be R_{,uv}=e^{s}\left(\frac{Q^{2}}{R^{\frac{3}{2}}}+\Lambda
R^{\frac{1}{2}}-\frac{1}{R^{\frac{1}{2}}} \right), \label{RRuv} \ee
\be s_{,uv}=e^{s}\left(-\frac{3Q^{2}}{2R^{\frac{5}{2}}}+
\frac{\Lambda}{2R^{\frac{1}{2}}}+\frac{1}{2R^{\frac{3}{2}}} \right).
\label{SSuv} \ee The constraint equations become \be
R_{,uu}=R_{,u}s_{,u}\;,\quad R_{,vv}=R_{,v}S_{,v}\, \label{constRs}.
\ee

In the next stage we simply omit the constraint equations
(\ref{constRs}) and keep the evolution equations
(\ref{RRuv},\ref{SSuv}) as our dynamical system. Recall that the
evolution equations form a closed hyperbolic system, which uniquely
determines the evolution of (properly-formulated) initial data. By
this we achieve several goals: First, a non-constrained dynamical
system is conceptually simpler to analyze than a constrained one;
Second, this allows us to explore and test our hypothesis that the
phenomenon of generic null-singularity formation inside
four-dimensional spinning BHs is essentially a property of the
evolution sector of the Einstein equations.
 \footnote
 {This is obviously a vague statement because the division of the Einstein equations into
 evolution and constraint subsystems is not unique, but depends on the choice of slicing.
 We expect, however, that this property of the Einstein equations will not be sensitive to
 the details of the foliation chosen. Note that in two effective dimensions the double-null
 formulation induces a unique division into evolution and constraint subsystems.}
 Finally, omitting the constraint equations retain the dynamics to the problem
 (it is the constraint sector which is responsible to properties like e.g. the Birkhoff theorem).
 This provides us with an effectively two-dimensional toy system aimed at mimicking
 dynamical properties of the Einstein equations in four dimensions.

 Next we recognize that the expression in the parentheses in Eq. (\ref{SSuv})
is the derivative of the expression in the parentheses in Eq.
(\ref{RRuv}) with respect to $R$. We can therefore write the two
equations as
\begin{equation}
R_{,uv}=e^{s}F(R)   \ \ \ ;\ \ \  s_{,uv}=e^{s}F'(R), \label{evf}
\end{equation}
where at this stage $F(R)$ denotes the specific function
\begin{equation}
F(R)=\frac{Q^{2}}{R^{\frac{3}{2}}}+\Lambda
R^{\frac{1}{2}}-\frac{1}{R^{\frac{1}{2}}},   \label{specificF}
\end{equation}
and hereafter a prime denotes a derivative
 of a function of one variable with respect to this variable.
These equations, which are semi-linear
nonhomogeneous wave equations, constitute the core of our model.

For later convenience we introduce another function, $h(R)$, defined
by its derivative:
\begin{equation}
F(R)=-h'(R).
\end{equation}
Note that $h(R)$ is defined up to an integration constant. The
semi-linear system now reads
\begin{equation}
R_{,uv}=-e^{s}h'(R)   \ \ \ ; \ \ \  s_{,uv}=-e^{s}h''(R),
\label{evh}
\end{equation}
 In the specific case (\ref{specificF}) we have
\begin{equation}
h(R)=2R^{\frac{1}{2}}+\frac{2Q^{2}}
{R^{\frac{1}{2}}}-\frac{2\Lambda}{3}R^{\frac{3}{2}} +const \ .
\label{speh}
\end{equation}
Note that this is $2r$ times the standard RNDS function
\begin{equation}
1-\frac{2m}{r}+\frac{Q^{2}}{r^2}-\frac{2}{3}\Lambda r^{2} \ ,
\end{equation}
where here the integration constant was expressed as $-4m$, $m$ being the ADM mass of
the corresponding static RNDS solution.

The final stage in constructing our toy system is to abandon the
specific function $h(R)$ of Eq. (\ref{speh}) and instead to explore
the semi-linear system (\ref{evh}) for a general function $h(R)$.
Again we define $F\equiv -dh/dR$, or
\begin{equation}
h(R)= - \int F(R)dR .
\label{hgen}
\end{equation}
This generalization is advantageous for several reasons. First, if indeed the
system (\ref{evh}) leads to generic null singularities, it is
plausible that this property will not be sensitive to the specific
functional form (\ref{speh}). Rather, we expect the qualitative
properties of our dynamical system to depend only on certain
qualitative features of $h(R)$. Note also that our primary goal is
to provide a simple toy model aimed at mimicking certain dynamical
features of e.g. the vacuum Einstein equations in four dimensions, and from this
perspective the spherically-symmetric electro-vacuum system of the
previous subsection should itself be regarded as a toy model; hence
there is no reason to firmly stick to the specific function
(\ref{speh}). Second, this extension of our view-point will allow us
to seek simple examples of functions $h(R)$ for which the general
solution of the system (\ref{evh}) may be constructed. Such solvable
examples would provide valuable insight into the dynamical
properties of this system.

\subsection{Application to two-dimensional black holes}

In addition to its role as a toy model for singularity formation,
our semi-linear system is also directly applicable to certain
dilatonic models of two-dimensional BHs. In the model developed by
Callan et al~\cite{CGHS} there is a dilaton $\phi (u,v)$, a
cosmological constant of the two-dimensional model $\lambda$, and
the metric is
\begin{equation}
ds^{2}=-e^{2 \rho (u,v)}dudv.
\end{equation}
The classical, matter-free, Einstein equations then yield two
evolution equations and two constraint equations. Transforming to
the new variables $R=e^{-2 \phi}$ and $S=2(\rho -\phi)$, the
constraint equations reduce to Eq. (\ref{constRs}), and the
evolution equations take the form (\ref{evh}), this time with the
generating function
\begin{equation}
h(R)=\lambda ^{2} (R+const). \label{hCGHS}
\end{equation}

This dilatonic two-dimensional model was later generalized to
include a Maxwell field as well as charged matter
fields~\cite{Ori2,thorlacius2,thorlacius}. Here, again, with the
same substitution $R=e^{-2 \phi},S=2(\rho -\phi)$ the classical
matter-free Einstein equations are reduced to Eqs.
(\ref{constRs},\ref{evh}) with the generating function
\begin{equation}
h(R)=\lambda ^{2} (R+Q^{2}/R+const), \label{chargedCGHS}
\end{equation}
where Q is a parameter proportional to the Maxwell field's charge.

\section{Basic mathematical properties}
In this section we introduce some basic features of our semi-linear
system (\ref{evf}).

\subsection{The gauge freedom}

Our semi-linear system (\ref{evf}) is invariant under a family of
gauge transformations. These are coordinate transformations which
preserve the double-null form of the metric: $u \rightarrow
\tilde{u}(u), v \rightarrow \tilde{v}(v)$.
 The variable $R$ is invariant under this coordinate transformation,
 but $s$ changes. Since $e^{s} \propto g_{uv}$,
it transforms like a covariant tensor of rank two, and one finds:
\begin{equation}
\tilde{s}=s-\ln\left(\frac{d\tilde{v}}{dv}\right)-
\ln\left(\frac{d\tilde{u}}{du}\right). \label{coo}
\end{equation}

The various quantities made of $R$ and $s$ may be classified
according to the way they transform under a gauge transformation.
The simplest are the {\it scalars}, namely quantities which are
unchanged. Obviously $R$ is a scalar. Apart from $R$ itself, there
is only one scalar made of  $R$ and $s$ and their first-order
derivatives: $e^{-s}R_{,u}R_{,v}$. Another useful, non-scalar,
quantity is $e^{-s}R_{,w}$, where hereafter $w$ stands for either
$u$ or $v$. This quantity is invariant to a transformation of $w$,
but not to transformation of the other null coordinate.

\subsection{The conserved fluxes}
\label{ev+cons} Consider the quantities \be \Phi \equiv
R_{,vv}-R_{,v}s_{,v} \ \ ; \ \ \ \Psi \equiv R_{,uu}-R_{,u}s_{,u} \
. \label{phipsi} \ee Differentiation $\Phi$ with respect to $u$, one
observes that
\be (R_{,uv})_{,v}-R_{,uv}s_{,v}-R_{,v}s_{,uv} \ee
identically vanishes by virtue of the field equations (\ref{evf}).
In a similar manner one finds that the derivative of $\Psi(u)$ with
respect to $v$ vanishes. Namely,
\begin{equation}
\Phi_{,u}=0   \ \ \ ; \ \ \  \Psi_{,v}=0 \ .
\label{conserv}
\end{equation}
 We shall refer to $\Phi(v)$ and $\Psi(u)$ as the two {\it conserved fluxes}
 (or simply {\it fluxes}). It is sometimes useful to express these fluxes as
 \begin{eqnarray}
\Phi(v)=e^{s}(e^{-s}R_{,v})_{,v}  \label{constraint v},  \\
\Psi(u)=e^{s}(e^{-s}R_{,u})_{,u}  \label{constraint u}.
\end{eqnarray}

One can easily verify that in a gauge transformation $u \rightarrow
\tilde{u}(u), v\rightarrow \tilde{v}(v)$ the two fluxes transform as
 \begin{eqnarray}
\tilde{\Phi}=\Phi (d\tilde{v}/dv)^{-2}
\\
\tilde{\Psi}=\Psi (d\tilde{u}/du)^{-2}
\end{eqnarray}
(namely like components of a covariant second-rank tensor).

Note that $\Phi(v)$ and $\Psi(u)$ are uniquely determined by the
initial data for $R$ and $s$ (this is most easily seen when the
characteristic initial-value formulation is used~\cite{second}).

{\it Application to the spherically-symmetric charged case:}

In the four-dimensional spherically-symmetric case, an important
problem is that of the RN solution perturbed by two fluxes of null
fluids, namely ingoing and outgoing fluxes. In this case, the dust
contribution to the energy-momentum tensor is~\cite{poisson}
 \be
T_{vv}^{dust}= L_{in}(v)/(4 \pi r^{2}) \ \ \ ; \ \ \ T_{uu}^{dust}=
L_{out}(u)/(4 \pi r^{2}) \ , \ee where $r$ is the area coordinate,
$u$ and $v$ are two null coordinates, and $L_{in},L_{out}$ denote
the ingoing and outgoing dust fluxes, respectively. {}From the
Einstein equations for $T_{vv}$ and $T_{uu}$ with the line element
(\ref{metric1}) one finds that \be L_{in}=r \left(  \frac{ f_{,v}
r_{,v}}{f}-r_{,vv} \right) \ \ \ ; \ \ \ L_{out}=r \left(  \frac{
f_{,u} r_{,u}}{f}-r_{,uu} \right) \ . \label{1cons1} \ee These
quantities are directly related to the conserved fluxes
$\Phi(v),\Psi(u)$ discussed above. In fact one can easily show,
using Eqs. (\ref{vara},\ref{phipsi}), that \be L_{in}=-\frac{1}{2}
\Phi(v) \ \ \ ; \ \ \ L_{out}=-\frac{1}{2} \Psi(u) \ . \ee We may
therefore regard the quantities $\Phi,\Psi$ as the generalization of
the spherically-symmetric null-fluid fluxes to arbitrary $h(R)$.

It is important to recall that in the spherically symmetric case
(\ref{specificF}) the semi-linear hyperbolic system (\ref{evf}) is
mathematically equivalent to the mass-inflation model~\cite{poisson}
with two arbitrary fluxes $L_{in}$ and $L_{out}$.

\subsection{The generalized mass function}

Consider the scalar quantity \be M(u,v) \equiv
e^{-s}\,R_{,u}\,R_{,v}+h_{0}(R) \ \label{Mdef} \ee where $h_{0}(R)$
is a certain member of the one-parameter family (\ref{hgen}) (namely
one associated with a certain choice of the integration constant).
This is a generalization of the mass parameter to dynamical cases
(see below). Note that for a given $F(R)$ the mass function is
defined up to an additive constant (associated with different
choices of $h_{0}(R)$).

One can easily show, using the field equation (\ref{evh}) for $R$,
that the derivatives of $M$ satisfy \be M_{,v}=e^{-s}\,R_{,u}
\Phi(v) \ \ ; \ \ \ M_{,u}=e^{-s}\,R_{,v} \Psi(u) \ . \label{Mdw}
\ee Also, differentiating the last equation with respect to $v$ and
recalling Eq. (\ref{constraint v}), one observes that $M$ satisfies
the simple field equation \be M_{,uv} =e^{-s}\Psi (u) \Phi (v) \ .
\label{Mduv} \ee

{}From Eq. (\ref{Mdw}) we see that when the fluxes $\Phi(v)$ and
$\Psi(u)$ vanish, the mass function becomes a fixed parameter. This
is the situation in the ``flux-free solution'' described below
(section \ref{vacuum}). Also when one flux vanishes (e.g. $\Psi$),
the mass function only depends on one null coordinate ($v$ in this
case), which is the situation in the generalized Vaidya solution
(discussed in the Appendix).

{\it Application to the spherically-symmetric charged case:}

As an illustration we consider here the mass function for
spherically symmetric charged black holes. In this case one
naturally defines $h_{0}$ by omitting the constant in Eq.
(\ref{speh}), namely
\begin{equation}
h_{0}(R)=2R^{\frac{1}{2}}+\frac{2Q^{2}}
{R^{\frac{1}{2}}}-\frac{2\Lambda}{3}R^{\frac{3}{2}}  \ .
\label{speh0}
\end{equation}
In terms of the original variables $r,f$ the mass function then reads
\begin{equation}
M(u,v)=2 r (1+\frac{Q^{2}}{r^{2}}-\frac{\Lambda}{3} r^{2}) +\frac{4
r r_{,v}r_{,u}}{f} \ . \ee

In the case of $\Lambda=0$, the last expression is just four times the function
$m(u,v)$ defined by Poisson and Israel~\cite{poisson}.
\footnote{In Ref.~\cite{poisson} $e^{2\sigma}$ is used instead of $f$.}
Rewriting Eqs. (\ref{Mdw}) and (\ref{Mduv}) in terms of $r,f,m,\sigma$
and the fluxes $L_{in},L_{out}$,
one recovers Eqs. (4.4) and (3.15)  therein for $m_{,v}$, $m_{,u}$, and $m_{,uv}$.

\section{The flux-free solution}
\label{vacuum} In this section we investigate a class of solutions
which is the generalization of the static RNDS family to general
$h(R)$. These are the solutions in which both $\Psi$ and $\Phi$
vanish. We first construct these solutions in Eddington-like
coordinates and then transform to Kruskal-like coordinates. We then
explore the singularities of these solutions.

\subsection{Construction in Eddington-like coordinates}

In the case considered here, \be \Psi=\Phi=0 \ , \label{fluxes0} \ee
Eqs. (\ref{constraint v},\ref{constraint u}) read \be
(e^{-s}R_{,v})_{,v}=0 \ \ \ ; \ \ \ (e^{-s}R_{,u})_{,u}=0 \ .
\label{fluxes02} \ee The first integral of these two equations is
\be R_{,v}=c_{u}(u) e^{s} \ \ \ ; \ \ \ R_{,u}=c_{v}(v) e^{s} \ .
\label{fluxes03} \ee The equation for $R_{,v}$ is invariant to a
transformation of $v$, but a transformation of $u$ affects the
function $c_{u}(u)$. However, the signs of $c_{u}$ is preserved in
such a gauge transformation, because we require the new null
coordinate $\tilde u$ to be future-directed, just like the original
$u$. The situation with the equation for $R_{,u}$ is exactly the
same (with the obvious interchange of $u$ and $v$). Thus, with the
aid of a gauge transformation we can bring both functions $c_{u}(u)$
and $c_{v}(v)$ to $\pm 1$, with the signs corresponding to those of
the original functions.

Consider first the case where $R$ is increasing with $v$ and
decreasing with $u$, namely $c_{v}>0$ and $c_{u}<0$ (this is
typically the situation outside a BH, though no further than the
cosmological horizon -- namely region I in Fig. \ref{fig3}).
\begin{figure}[htb]
 \vspace{0.5cm}
\epsfxsize=0.4\textwidth \centerline{\epsffile{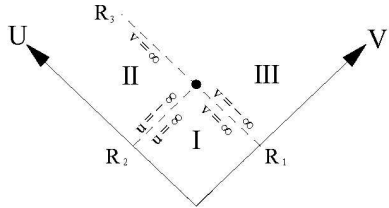}}
\vspace*{0.3cm} \caption{\small \bf \sl Regions $I$,$II$ and $III$
in the flux-free solution.} \label{fig3}
\end{figure}
Then the gauge transformation described above leads to
\be
R_{,v}=e^{s}=-R_{,u} \ . \label{fluxes01} \ee
This implies that both
$R$ and $s$ are functions of a single variable, \be x=v-u \ ,  \ee
and these two functions satisfy \be R_{,x}=e^{s}   \ . \ee
Then the field equation (\ref{evh}) for $R_{,uv}$ reads
\be R_{,xx}=R_{,x}h'(R)=\frac{d}{d x} h(R(x)) \ ,
\label{Rxx} \ee
yielding the first order ODE
\be
 R_{,x}= h(R)  .
\label{Rdx} \ee Thus, $R(x)$ is given by its inverse function
 \footnote{The full integral of Eq. (\ref{Rxx})
obviously involves two integration constants. The first one is
already embodied in the definition of $h$ in Eq. (\ref{Rdx}). The
second one is an arbitrary constant to be added to the right-hand
side of Eq. (\ref{solutionr}). This constant has no physical
meaning, however, as it may be absorbed by a gauge transformation
 e.g. $v \to v+const$, which merely shifts $x$ by a constant.}
\be
 x(R)=\int^{R} \frac{1}{h(R')} dR' \ .
\label{solutionr} \ee
Then $s(x)$ is given by \be
 s=ln(h(R)) \ .
\label{solutions} \ee The second field equation, namely Eq.
(\ref{evh}) for $s_{,uv}$, is automatically satisfied, as one can
easily verify. Note that this solution is {\it static}, in the sense
that it only depends on the spatial variable $x=v-u$. Note also from
Eqs. (\ref{Rdx}) or (\ref{solutions}) that $h$ must be positive in
this case.

Next let us consider the case where $R$ is decreasing with both $v$
and $u$, which is typically the situation inside a BH (region II in
Fig. \ref{fig3}). Then instead of Eq. (\ref{fluxes01}) we now get
\be R_{,v}=-e^{s}=R_{,u} \ . \label{fluxes04} \ee Correspondingly we
now define \be x=v+u \ ,  \ee and both $R$ and $s$ are functions of
$x$ only, satisfying \be R_{,x}=-e^{s}   \  \ee this time.
Substituting again in the field equations (\ref{evh}), one finds
that Eqs. (\ref{Rdx}) and Eq. (\ref{solutionr}) for $x(R)$ still
hold, but there is a sign change in the expression for $s$, namely
$s=ln(-h(R))$. Note that $h$ is negative in this case. The general
expression for $s$ which holds in both cases is obviously \be
 s=ln(|h|) \ .
\label{solutionsg} \ee

Thus, the flux-free solution comes in two versions: The
``external-type'' version, which depends on the spatial variable
$v-u$, and the ``internal-type'' version, which depends on the
temporal variable $v+u$. The first version occurs in regions where
$h>0$, and the second occurs when $h<0$ (typically inside a BH).
Equations (\ref{solutionr}) and (\ref{solutionsg}) hold in both
cases.

In its both versions, the flux-free solution is in fact a
one-parameter family of solutions (for given $F(R)$), due to the
arbitrary integration constant in the definition of $h$ in Eq.
(\ref{hgen}). The situation is somewhat confusing because to a
specific flux-free solution a function $h(R)$ is associated in two
different ways: (i) through the field equation (\ref{evh}) (to be
satisfied by the solution), and (ii) through its explicit
construction via Eq. (\ref{solutionr}). It must be noticed that
$h(R)$ appears in two conceptually different ways in these two
occasions: In the field equation it appears as an {\it equivalence
class} (because the field equations only depend on $dh/dR$), and in
the construction procedure it appears as a {\it one-parameter family
of distinct functions}. To clarify this notational confusion we
reserve $h(R)$ of Eq. (\ref{hgen}) to denote the equivalence class,
and introduce the notation \be H(R)=h_{0}(R)+c   \ ,  \label{HR} \ee
where $h_{0}(R)$ is a representative of the equivalence class $h(R)$
(already mentioned above), and $c$ is an arbitrary constant which
distinguishes the various members of this equivalence class. Thus,
we rewrite the flux-free solution as
 \be
s=ln(|H(R(x))|)  \ \ \ ; \ \ \ \frac{dR}{dx}=H(R)  \ .
\label{solutionH} \ee (with $x=v \pm u$ as above), namely
 \be
x(R)=\int^{R} \frac{1}{H(R')} dR'
 \label{solutionHR} \ee
The parameter $c$ may be interpreted as one associated with the system's mass.
Indeed, the mass function (\ref{Mdef}) reads for flux-free solutions
\be
M(u,v)=h_{0}(R)-H(R)=-c \ .
 \label{fluxfreem} \ee
In the spherically-symmetric four-dimensional case, the flux-free
solution is just the RNDS family of solutions, whose ADM mass is
related $c$ through $m=-c/4$. We therefore write $H$ in this case as
\begin{equation}
H(R)=2R^{\frac{1}{2}}+\frac{2Q^{2}}
{R^{\frac{1}{2}}}-\frac{2\Lambda}{3}R^{\frac{3}{2}} -4m  \ .
\label{HRNDS}
\end{equation}

\subsection{Horizons}

The solutions constructed above become pathological at any value
$R=R_{0}$ for which $H$ vanishes. {}From Eq. (\ref{solutionH}) $s$
diverges there to $-\infty$. Also Eq. (\ref{solutionHR}) implies
(assuming finite $F(R_{0})$) that $x$ diverges at $R=R_{0}$, meaning
that either $u$ or $v$ is unbounded there. This phenomenon is
analogous to the coordinate singularity at the horizon of the
Schwarzschild solution, when the metric is expressed in double-null
Eddington coordinates. In our case, too, this coordinate singularity
may be overcome by transforming to new, Kruskal-like, null
coordinates, as shown in the next subsection. To this end, however,
we must first analyze the asymptotic behavior of $s$ and $x$ on
approaching the horizon in the original Eddington-like gauge.

We define (for each horizon): \be K \equiv  H'(R_{0}) \ee and assume
$K \neq 0$, therefore \be H(R) \simeq K(R-R_{0}) \ee in the
neighborhood of the horizon. Note that for $K>0$, $dx/dR=1/H$ is
negative for $R<R_{0}$ and positive for $R>R_{0}$, hence $x \to -
\infty$ at both sides of the horizon. Similarly, for $K<0$ at both
sides $x \to + \infty$. Therefore $Kx \to - \infty$ for both $K>0$
and $K<0$, and at both sides of the horizon.

In the horizon's neighborhood Eq. (\ref{solutionH}) reads \be R_{,x}
\simeq K(R-R_{0}), \label{nearhRdx} \ee yielding \footnote{In the
general solution to Eq. (\ref{nearhRdx}) the right-hand side of Eq.
(\ref{nearhorizonR}) should be multiplied by an arbitrary constant.
However, this constant may be omitted with no loss of generality
because it can be absorbed by a shift in $x$. Such a shift merely
corresponds to a gauge transformation $v \to v+const$ as noted
above.} \be R-R_{0} \simeq \pm e^{K x} \  . \label{nearhorizonR} \ee
Also the same equation for $s$ reads \be
 s=ln(|H|) \simeq K x +ln(|K|) \ .
\label{solutionsh} \ee Thus, both $s$ and $x$ diverge
logarithmically in $R-R_{0}$.

The occurrence of sign flips at the horizon in some of the above
expressions complicates the analysis. To help clarifying this
confusion we define, for each horizon, the quantity
\[
X(R)\equiv sign(R-R_{0}) e^{K x} \ .
\label{Xdef}
\]
$X$ (unlike $x$) is continuous and monotonous across the horizon,
and it vanishes at the horizon itself. The comparison of Eqs.
(\ref{nearhorizonR}) and the definition of $X$ immediately yields
$dR/dX=\pm 1$ at the horizon, and a closer look at the signs reveals
that at both sides \be \frac {dR}{dX}=1 \ . \label{RdX0} \ee

Since $dX/dx=KX$, $R(X)$ satisfies the same differential equation at
both sides: \be \frac {dR}{dX}=\frac {H(R)}{K X} \ . \label{RdX} \ee
This, combined with Eq. (\ref{RdX0}), implies that $X(R)$ is
analytic across the horizon (provided that $H(R)$ itself is analytic
in a neighborhood of $R=R_{0}$, which we assume). Note that $H/X$ is
analytic too, and it gets the non-vanishing value \be \frac {H}{X}=K
\ \label{hX} \ee at the horizon.

We shall primarily be interested in functions $H(R)$ admitting (at
least) three simple roots $R_{i}$ ($i=1,2,3$), ordered $R_{3}<R_{2}<R_{1}$,
such that $H$ is positive at $R_{2}<R<R_{1}$ and negative at
$R_{3}<R<R_{2}$, as shown in Fig. \ref{fig1}. An archetype is the
function $h(R)$ corresponding to the spherically-symmetric
electro-vacuum solutions, Eq. (\ref{HRNDS}), which (for sufficiently
small $Q$ and $\Lambda$, and sufficiently large $m$)
admits three roots. These roots correspond to the cosmological horizon ($R_{1}$),
the event horizon ($R_{2}$), and the inner horizon ($R_{3}$), as shown in Fig.
\ref{fig3}.
\begin{figure}[htb]
\vspace{0.5cm} \epsfxsize=0.8\textwidth
\centerline{\epsffile{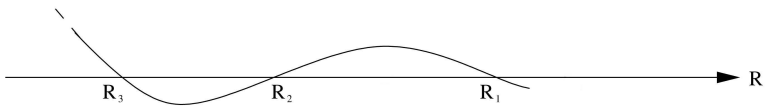}} \vspace*{0.3cm} \caption{\small
\bf \sl The function $H(R)$, which has three roots: $R_{1}$, $R_{2}$
and $R_{3}$.} \label{fig1}
\end{figure}

The horizons divide the spacetime into three regions which we denote
I,II,III, as shown in Fig. \ref{fig3}. We shall primarily be
concerned here with the regions I and II (region III will not
concern us here, except at the very neighborhood of the cosmological
horizon). Note that $H$ is positive in region I and negative in
regions II and III.

As was shown above, in the Eddington gauge $s$ diverges on the three
horizons. The divergence at the IH does not pose any difficulty---in
fact investigating this divergence and its physical implications is
one of our primary goals. However, the divergence of $s$ on the
event and cosmological horizons does pose an undesired feature: We
would like to explore a situation in which a (flux-free) solution of
the type described above, which includes the three horizons, emerges
from regular initial data prescribed on some compact spacelike
initial hypersurface $\Sigma_{0}$. Furthermore we want the horizons'
intersection point to be included in $D_{+}(\Sigma_{0})$. To this
end, $\Sigma_{0}$ must intersect both the event and cosmological
horizons (but {\it not} the inner horizon). The divergence of $s$
(and also $u$ or $v$) on these two horizons renders the Eddington
gauge inappropriate for such a regular initial-value set-up. We
shall therefore proceed now to transform the Eddington coordinates
into Kruskal-like coordinates with respect to the event and
cosmological horizons.

\subsection{Transforming to Kruskal-like coordinates}

The construction of the Kruskal-like coordinates in our case is
similar to the standard procedure in e.g. the Schwarzschild
spacetime---except that here we ``Kruskalize'' $u$ with respect to
the EH and $v$ with respect to the cosmological horizon.

Let us define on each horizon $R=R_{i}$:
\[k_{i} \equiv \mid\! H'(R_{i})\!\mid \ \ (i=1,..,3) \]
(namely, it is the $|K|$ value associated with the $i$'th horizon.)
\footnote{In the RNDS case, namely the function $H(R)$ of Eq.
(\ref{HRNDS}), $k_{1,2,3}$ correspond to twice the quantities
$\kappa_{co,ev,in}$ mentioned in the Introduction. } Consider first
the EH, $R=R_{2}$. Here $K>0$, hence $x$ diverges to $- \infty$. In
both regions I and II the Eddington coordinate $v$ is regular along
the EH but $u$ diverges (see Fig. \ref{fig3}). In region I $H>0$,
hence $x=v-u$, and the divergence of $x$ means that $u\to +\infty$.
On the other hand in region II $H<0$, hence $x=v+u$, and the
divergence of $x$ now implies that $u\to -\infty$. Correspondingly
we define \be U=-e^{-k_{2}\,u} \ee in region I, and \be
U=e^{k_{2}\,u} \ee in region II. Then U is continuous across the EH,
and is monotonously-increasing (namely future-directed) everywhere;
it is negative at region I, positive at region II, and vanishes at
the EH. This transformation cures the divergence of $s$, as we show
below.

Next we consider the cosmological horizon, $R=R_{1}$. Here $K<0$,
hence $x$ diverges to $+ \infty$, meaning that $v\to +\infty$ on
approaching the horizon from region I, whereas $u$ is regular.
\footnote{The detailed behavior of the various functions in Region
III will not really concern us here. The construction below
guarantees that in the Kruskal gauge all the relevant functions are
analytic across the cosmological horizon, and it will be sufficient
for us to define these functions through an analytic extension from
region I to region III.} We thus define (in region I) \be
V=-e^{-k_{1}\,v} \ . \ee Again, $V$ is a future-directed null
coordinate which takes negative values in region I and vanishes at
the cosmological horizon. Since $V$ is continuous across the EH, it
takes negative values in region II as well.

The variable $R$ is invariant under the coordinate transformation
($u\to U,v\to V$). Therefore, $R$ is formally given as a function of
$U$ and $V$ through \be R(U,V)=R(x(U,V)) \ , \label{kruskalR} \ee
where $x(U,V)=v(V)\pm u(U)$, and $R(x)$ is defined through its
inverse function (\ref{solutionHR}). On the other hand $s$ is
modified in the gauge transformation according to Eq. (\ref{coo}).
We shall denote our new Kruskal $s$ by $S$. It satisfies \be
e^{S}=e^{s} \frac{du}{dU}\frac{dv}{dV} \ . \ee In both regions I and
II we have $dV/dv=-k_{1}V$, $dU/du=\pm k_{2}U$, and $e^{s}=\pm H$.
The signs properly combine to yield \be e^{S}=
\frac{H(R)}{k_{1}k_{2}UV} \label{kruskalS} \ee in both regions.
Equations (\ref{kruskalR}) and (\ref{kruskalS}) constitute the
flux-free solution in the Kruskal-like gauge.

We now proceed to show the regularity of $R$ and $S$, and their
smoothness as functions of $U$ and $V$, at both the event and
cosmological horizons. Considering the EH first, we denote by
$X_{2}$ the function $X(R)$ (defined above) associated with the EH.
Noting that at the EH $K>0$ and hence $k_{2}=K$, we write \be X_{2}
\equiv sign(R-R_{2}) e^{k_{2} x} \ . \label{X2} \ee One finds
(treating carefully the flipping signs) that \be X_{2} =-U
(-V)^{-k_{21}} \ , \label{X2UV} \ee where hereafter $k_{ij} \equiv
k_{i}/k_{j}>0$ for any $i,j=1...3$. Since $V$ is strictly negative
along the EH, and $R$ is an analytic function of $X_{2}$ (as
establish above for a general horizon), we conclude that $R(U,V)$ is
analytic in the neighborhood of the EH. To analyze the variable $S$,
we note that
\[
U V=X_{2} (-V)^{1+k_{21}} \ ,
\]
therefore \be e^{S}= \frac{1}{k_{1}k_{2}}\frac{H(R)}{X_{2}}
(-V)^{-(1+k_{21})} \ . \label{SEH} \ee
As was established in the
previous subsection, $H/X_{2}$ is a regular function of $R$ (or
$X_{2}$) which takes the value $K=k_{2}>0$ at the EH. Therefore
$S(U,V)$ too is a regular (in fact analytic) function of $U$ and $V$.

The cosmological horizon is treated in an analogous manner, except
that here we do not need to explicitly analyze the various functions
in region III: Instead we simply extend the relevant functions
analytically from region I into region III. Note that here $K<0$ and
therefor $k_{1}=-K$. Correspondingly we find (for region I) \be
X_{1}\equiv sign(R-R_{1}) e^{-k_{1} x}=-e^{k_{1} (u-v)} \ .
\label{X1} \ee This yields \be X_{1}=V (-U)^{-k_{12}} \ .
\label{X1UV} \ee Again we see that $R(U,V)$ is analytic in the
neighborhood of the cosmological horizon, because $R(X_{1})$ is
analytic (and $U$ is strictly negative). Recalling that
\[
U V=-X_{1} (-U)^{1+k_{12}} \ ,
\]
we obtain \be e^{S}= \frac{1}{k_{1}k_{2}}\frac{-H(R)}{X_{1}}
(-U)^{-(1+k_{12})} \ . \label{Scosmo} \ee Again,  $-h/X_{1}$ is a
regular function of $R$ (or $X_{1}$) which takes the value
$-K=k_{1}>0$ at the cosmological horizon. Therefore we conclude
again that both $R(U,V)$ and $S(U,V)$ are analytic functions in the
neighborhood of the cosmological horizon. Then in region III
$R(U,V)$ and $S(U,V)$ are {\it defined} to be the analytic extension
of the corresponding functions from region I across the cosmological
horizon.

In fact it is straightforward to show that $R(U,V)$ and $S(U,V)$ are regular at $U,V<0$
not only in the neighborhood of the event and cosmological
horizons, but also in the entire range $R_{2}\geq R \geq R_{1}$ and its neighborhood
(provided that $H(R)$ itself is regular throughout).
Furthermore, for any smooth spacelike initial hypersurface $\Sigma_{0}$ which intersects
the event and cosmological horizons,
the initial data for $R$ and $S$ (corresponding to the
flux-free solution in the $U,V$ coordinates) are regular throughout.
\footnote{As we show below the solution typically develops a singularity
at $R = R_{3}$, and the same
applies to roots $R_{0} > R_{1}$ of $H(R)$ if such additional roots exist.
Nevertheless the spacelike hypersurface $\Sigma_{0}$ cannot intersect $R = R_{3}$
or $R = R_{0}$.}

 \section{Singularities in the flux-free solution}
In this section we analyze the singularities that appear in the
flux-free solution using the expressions that were derived in the previous section.
As before, we shall assume that
$H(R)$ has three simple roots at $R_{1}$, $R_{2}$ and $R_{3}$, with $H$ positive
at $R_{2}<R<R_{1}$ and negative at $R_{3}<R<R_{2}$
(additional roots at $R>R_{1}$ and/or $R<R_{3}$ are allowed; see last subsection).
We divide the discussion into two types of
singularities  --- the vertex singularity and the IH singularity.

\subsection*{The vertex singularity}
In the Kruskal-type coordinates defined in the previous section
$(U,V)=(0,0)$ is the intersection point of the three horizons (see
Fig. \ref{fig3}). At this point $R$ is many-valued
(for example it admits a fixed value $R=R_{i}$ along the i'th horizon).
Furthermore,
\[S \to + \infty\]
as $U,V \to 0$.
This divergence proceeds in a slightly different manner along
various paths towards the vertex $(0,0)$. For example,
along curves of constant $R\neq R_{i}$, $S\propto \log|U|+\log|V|$,
as can be seen from Eq. (\ref{kruskalS}).
Also, Eqs. (\ref{SEH}) and (\ref{Scosmo}) imply that
\be e^{S}=\frac{1}{k_{1}}\,(-V)^{-(1+k_{21})} \label{Sehor} \ee
along the event horizon, and
\be e^{S}=\frac{1}{k_{2}}\,(-U)^{-(1+k_{12})} \label{Schor} \ee
along the cosmological horizon. Since $k_{12}>0$, $S \to  +\infty$
along both horizons on approaching $U=V=0$.

\subsection*{The singularity at the inner horizon}

Next we analyze the asymptotic behavior of the flux-free solution on approaching the
IH from region II, namely the limit $V \to 0_{-}$ in the range $U>0$.
To this end we use the same
Kruskal-like coordinates $(U,V)$ introduced above, which allow for regular
flux-free initial data on $\Sigma_{0}$.

 {}From Eq. (\ref{RdX}) applied to the event horizon we deduce that
 $R$ decreases throughout region II, hence $R\ge R_{3}$.
 We denote by $X_{3}$ the function $X(R)$
(defined in the previous section) associated with the inner horizon.
Noting that at the IH $K<0$ and hence $k_{3}=-K$, we write for
region II (recalling $x=v+u$) \be X_{3} \equiv
e^{-k_{3}\,x}=(-V)^{k_{31}}\,U^{-k_{32}} \ .  \ee Substitution in
Eq. (\ref{kruskalS}) (which is valid in region II as well) yields
\be
e^{S}=[\frac{1}{k_{1}k_{2}}\frac{-H(R)}{X_{3}}\,U^{-(1+k_{32})}]\,
(-V)^{k_{31}-1}. \label{SIH} \ee {}From Eq. (\ref{hX}) we obtain
along the IH $H(R)/X_{3}=-k_{3}$. Therefore, for any curve in region
II which approaches the IH (at some $U>0$), the term in squared
brackets approaches a finite value as $V \to 0$. Along such a curve
$e^{S}$ behaves as \be e^{S} \approx \(
\frac{k_{3}}{k_{1}k_{2}}\,U^{-(1+k_{32})} \) \, (-V)^{k_{31}-1}.
\label{SIH1} \ee We find that $S \to \infty$ if $k_{1}>k_{3}$, $S
\to -\infty$ if $k_{1}<k_{3}$ and $S$ is regular in the case of
$k_{1}=k_{3}$.

The singularity at the IH is locally gauge-removable; Namely, it may be
removed by transforming from $V$ to a new Kruskal-type coordinate
\[ V_{3}\equiv -e^{-k_{3}\,v} =-(-V)^{k_{31}} \ , \]
defined with respect to the IH. One then obtains in the $(U,V_{3})$
coordinates, in full analogy with Eqs. (\ref{Scosmo}) and
(\ref{Schor}), \be e^{S_{3}}=
\frac{1}{k_{2}k_{3}}\frac{-H(R)}{X_{3}} U^{-(1+k_{32})} \ .
\label{Sinhor} \ee which admits the regular limit \be
e^{S_{3}}=\frac{1}{k_{2}}\,U^{-(1+k_{32})} \label{Sinh} \ee at the
IH. However, this transformation will destroy the regularity of $S$
on the cosmological horizon. In particular it will spoil the
regularity of the initial data on $\Sigma_{0}$ as the latter
intersects the cosmological horizon. Hence the divergence of $S$ at
the IH (for $k_{1}\neq k_{3}$) is globally non-removable. {}From the
perspective of the finite-time blow-up phenomenon which concerns us
here, it is mandatory to introduce regular initial data, and the
divergence of $S$ on the IH is inevitable.

We now introduce three additional quantities and explore their
asymptotic behavior at the IH. These quantities will serve as gauge-invariant
indicators for certain aspects of regularity or irregularity at the IH.

i) {\bf $e^{-S}\,R_{,V}\,R_{,U}$ :} This quantity is interesting because (unlike $S$)
it is a {\it scalar}, namely invariant under a gauge transformation.
We can therefore calculate it directly from the original Eddington-like gauge through
Eq. (\ref{solutionH}):
\[ e^{-S}\,R_{,V}\,R_{,U}=e^{-s}\,(R_{,x})^{2}=|H|  \]
which actually vanishes on the IH.

(ii) {\bf $e^{-S}\,R_{,V}$ :} This quantity is not a scalar, yet it
is invariant under a transformation of $V$. In a gauge
transformation $u \to \tilde{u}, v \to \tilde{v}$ it is simply
multiplied by $d \tilde{u}/du$. We restrict attention here to
transformations $u \to \tilde{u}$ which are regular in the interior
of region II (i.e. at $U>0$). We then conclude that although the
actual value of  $e^{-s}R_{,v}$ is modified, the divergence of this
quantity at the IH is invariant to a gauge transformation. In the
Eddington gauge Eq. (\ref{solutionH}) implies $e^{-s}R_{,v}=-1$. In
the Kruskal-like gauge we obtain
\[ e^{-S}\,R_{,V}=-dU/du=-k_{2}U \ .  \]

(iii) {\bf $S_{,U}$ :}  This quantity too is invariant under a
transformation $v \to \tilde{v}$, because the term $-ln(d
\tilde{v}/dv)$ in Eq. (\ref{coo}) is independent of $U$. Therefore
its regularity at the IH is gauge invariant. We calculate this
quantity in the Eddington gauge:
\[ s_{,u}=H_{,u}/H=\frac{dH}{dR}R_{,x}x_{,u}/H=\frac{dH}{dR}=-F(R) \ ,  \]
which is presumably regular.

The regularity of $S_{,U}$ (or, more generally, of its
gauge-transformed counterpart $\tilde{s}_{,\tilde{u}}$) is a useful
indicator for whether the divergence of $S$ at $V=0$ is locally
gauge-removable or not. To show this, we choose in region II a line
$U=U_{0}>0$ which intersects the IH. We then define the new
coordinate $\tilde{v}$ to be the affine parameter along this line,
setting e.g. $\tilde{v}=0$ at the IH. Then $\tilde{s}=const \equiv
\tilde{s}_{0}$ everywhere along this line, and in particular at
$\tilde{v} \to 0$. Now, at any $U>0$ we have
\[ \tilde{s}(U,\tilde{v})=
\tilde{s}_{0}+\int_{U_{0}}^{U}{\tilde{s}_{,u}du} \ .  \]
Therefore regularity of $\tilde{s}_{,\tilde{u}}$ implies regularity of $\tilde{s}$
at the IH, and vice versa.
\footnote {Obviously in the flux-free solutions the local gauge removal of the divergence
of $S$ can be demonstrated explicitly, see Eq. (\ref{Sinh}). Yet the
indicator $S_{,U}$ is useful for the wider class of flux-carrying solutions.}

On the other hand, the divergence of $e^{-S}R_{,V}R_{,U}$ and
$e^{-S}R_{,V}$ indicates (qualitatively speaking) to what extent the
gauge-invariant variable $R$ is smooth at the IH. Recall that after
the divergence of $S$ has been removed (locally) by a gauge
transformation $V \to V_{3}$, the solution $(R,S_{3})$ is continuous
at the IH, so we are left with the issue of the next-level
regularity, namely $C^{1}$ smoothness at the IH. The quantities
$e^{-S}R_{,V}R_{,U}$ and $e^{-S}R_{,V}$ both provide information on
this issue, in a gauge-invariant manner. \footnote {We consider here
both quantities (i,ii), even though both essentially probe the
smoothness of $R$ at the IH, because each of them has its own
advantages and disadvantages. The quantity (i) is advantageous
because it is a scalar. On the other hand in certain cases
$e^{-S}R_{,V}$ diverges, indicating the non-smoothness of $R$, and
yet $e^{-S}R_{,V}R_{,U}$ vanishes. This happens because
(qualitatively speaking) although the $V$-derivative of $R$
diverges, its $U$-derivative vanishes. This situation typically
happens in Vaidya-like solutions where only ingoing flux is present.
In this sense the indicator $e^{-S}R_{,V}$ is more robust.}

In the flux-free solution all the above indicators (i-iii) are
regular (even though $S$ itself diverges). As we shall discuss in
the next section, this situation presumably changes for indicators
(i) and (ii) in the more general flux-carrying solutions.

The discussion above was based on the Kruskal-like coordinates $U,V$.
It should be emphasized, though, that the singularity structure
is in fact {\it gauge-invariant}, in the following sense: For any new coordinates
$\tilde{U}(U),\tilde{V}(V)$ related to $U,V$ in a regular manner,
\footnote{by this we also mean that $\tilde{U}(U)$ is monotonic with
non-vanishing derivative, and the same for $\tilde{V}(V)$.}
the structure
of both the vertex and the IH singularities will be the same as in the original coordinates.
In other words, any choice of null coordinates $\tilde{U},\tilde{V}$ for which the
flux-free initial data are regular on $\Sigma_{0}$ and its immediate neighborhood,
will lead to the same structure of singularity.
(Though, obviously the vertex singularity will shift from (0,0) to some other point
$\tilde{U}_{0},\tilde{V}_{0}$.)

\subsection*{The relation between the PDE
and general-relativistic notions of singularity}

The issue which primarily concerns us in this paper is that of
finite-time blowup in a non-linear hyperbolic system of PDEs
mimicking the Einstein equations. To this end we throughout adopt
the standard terminology of PDEs in considering the onset of
irregularities (``singularities'') in the evolving solutions. This
terminology is quite different from that used in General relativity
(GR) for discussing spacetime singularities. In this subsection we
discuss how the various irregularity phenomena admitted by our PDE
toy system are interpreted from the GR point of view.

We have seen before that the three-roots flux-free solutions
typically develop two types of singular phenomena: the vertex
singularity and the IH singularity. It should thus be emphasized
that both phenomena do {\it not} imply a General-relativistic
spacetime singularity, as we now discuss.

The divergence of $S$ at the vertex
$U=V=0$ fails to mark a General-Relativistic singularity because
(in a typical application of our system
to a general-relativistic spacetime) no divergence of curvature is involved.
(Note that $S$ is a gauge-dependent quantity, and gauge-invariant quantities like
$e^{-S}R_{,V}R_{,U}$ do not diverge at the vertex.)
The divergence of the metric function $g_{uv} \propto e^{S}$ simply implies that
the proper-time distances between the vertex and points in region I are infinite---
namely, the vertex becomes {\it future timelike infinity}.

Similarly, the divergence of $S$ at the IH
fails to mark a spacetime singularity because
it may be locally removed by a gauge transformation
$V \to V_{3}$. Smoothness indicators like $e^{-S}R_{,V}R_{,U}$
or $e^{-S}R_{,V}$ (and others not mentioned here) all indicate that subsequent to
such a coordinate transformation the manifold is perfectly regular (in fact analytic) at the IH.

Despite the failure of the vertex and IH singularities to mark genuine GR singularities, both
phenomena actually bear crucial implications to the structure and features of the
GR black-hole spacetime, as we now discuss.

The vertex singularity functions as timelike infinity, as we already mentioned.
In fact it {\it defines} the event and cosmological horizons---both are null lines which
intersect the vertex singularity at their future.
\footnote{In the flux-free solutions the horizons may alternatively be defined
as null lines of
constant $R$, or as the locus of $H=0$. These definitions do not hold in the
more general flux-carrying solutions.}
Furthermore, from the standard hyperbolic-PDE point of view the occurrence of a
singularity at the vertex implies the presence of a Cauchy horizon in the $(U,V)$
manifold at $V=0$. This property carries over to the GR terminology---
the inner horizon at $V=0$ is indeed a Cauchy horizon of the spacetime.
Thus, the very presence of a Cauchy horizon in e.g. the RNDS family of black
holes may be attributed to the divergence of $S$ at the vertex.

Next let us consider the General-relativistic implications of the IH
singularity. As was mentioned above the divergence of $S$ at $V=0$
may be gauge-removed in a local manner only: Curing the divergent
$S$ at the IH will inevitably lead to blow-up of $S$ at the
cosmological horizon. The mismatch of $S$ between the inner and
cosmological horizons is an inherent, gauge-invariant, global
feature of the typical three-root flux-free solutions. In the
general-relativistic language this is translated into {\it infinite
blue-shift} which takes place at the IH of the RNDS black holes (as
well as their rotating counterparts). \footnote{This holds in the
case $k_{3}>k_{1}$. In the other case $k_{3}<k_{1}$ there is an
infinite {\it red-shift} at the IH.} This divergent blue-shift, in
turn, leads to a null curvature singularity at the IH when the BH is
generically perturbed. But this relativistic inner-horizon
singularity is represented in our PDE terminology by the divergence
of quantities like $e^{-S}R_{,V}R_{,U}$ and $e^{-S}R_{,V}$ (which,
as was mentioned above, only diverge when fluxes are added).

\subsection*{Three roots versus four roots}

The analysis and discussion so far was independent of the behavior
of $H(R)$ at $R>R_{1}$. The extensions of the spacetime diagram in
Figs 1 or 3 to the right of the cosmological horizon will depend the
properties of $H(R>R_{1})$. We may conceive of several options:

(i) $H(R)$ admits an additional root at $R_{0}>R_{1}$;

(ii) No additional root exists at $R>R_{1}$; either
(iia) $H(R)$ diverges (or admits an irregularity) at a certain finite $R_{s}>R_{1}$; or
(iib) $H(R)$ is regular---and negative---for any finite $R>R_{1}$
(the asymptotic behavior of this function as $R \to \infty$ will not concern us here).

Consider first the case (i). In this case the dynamics in region III
is essentially the same as in region II. In particular, the line
$H_{3}$ in Fig. 3 (the IH) which borders region II has a counterpart
at the future boundary of region III---namely the null line located
at $U=0,V>0$, which we denote $H_{0}$. The behavior of $R$ and $S$
on approaching $H_{0}$ just reflects that at $H_{3}$, with the
obvious interchange of the horizons' indices $3 \to 0$ and $1 \to
2,2 \to 1$. In particular $R$ gets the fixed value $R_{0}$, whereas
$S$ diverges. The sign of divergence is $-\infty$ (implying infinite
blue-shift) for $k_{0}>k_{2}$ and $+\infty$ (infinite red-shift) for
$k_{0}<k_{2}$. Essentially the spacetime diagram of Fig. 3 (with the
$H_{0}$ singularity added) would exhibit a reflection symmetry with
respect to a vertical line which passes through the vertex at
$U=V=0$. The singularity will have a symmetric ``V-shape'' in this
case, namely two null singularities which emerge from a common
vertex singularity.

Note that additional roots at $R>R_{0}$ will not affect the
spacetime diagram because we terminate the solution at the $H_{0}$
singularity, namely at $R=R_{0}$. For the same reason, additional
roots at $R<R_{3}$ will not affect the spacetime diagram of Fig. 3
(regardless of whether a root $R_{0}>R_{1}$ exists or not).

Turn now to discuss the case (ii), namely $H(R)$ has no roots at $R>R_{1}$.
In this case region III will terminate at a spacelike boundary line, corresponding to
$R=R_{s}$ in case (iia) or $R \to \infty$ in case (iib). This spacelike line,
which intersects the vertex $U=V=0$,
will constitute a singularity of our PDE system. \footnote{ $S$ diverges in case (iia),
and $R$ (and possibly also $S$) diverges in case (iib).}
Recall, however, that in the GR terminology
the spacelike boundary will not necessarily be a spacetime singularity; Instead it may
well mark a regular boundary at timelike infinity.

The function $H(R)$ of Eq. (\ref{HRNDS}) corresponds to the RNDS black-hole
solutions. In this case there are only three roots, corresponding to the cosmological,
event, and inner horizons. This function belongs to the class (iib), and the future
spacelike boundary of region III marks the future timelike infinity of the
external de Sitter universe.

\section{The conjecture about the singularity formation}

The semi-linear system (\ref{evh}) is the main objective of this
paper and the subsequent paper~\cite{second}. In the previous
sections a class of special solutions was constructed which
demonstrate a finite time blow-up. We saw various features of the
singularity which evolved from regular initial data. The solutions
that we constructed were flux-free, namely ``static''. They all
correspond to flux-free ($\Psi=\Phi = 0$) initial data. This leaves
us with the following open question: How will these solutions change
if we perturb them by modifying the initial data, such that
($\Psi,\Phi \neq 0$)? We are particularly interested in the
finite-time blowup phenomenon and the structure of the singularity
exhibited by the flux-free solutions: To what extent it is stable to
the perturbations?

To be more specific, and to set the notation for the discussion
below, let us consider a specific generating function $h(R)$ and a
specific three-root (or more) flux-free solution $H(R)=h_{0}(R)+c$.
The flux-free initial data are prescribed on an initial hypersurface
$\Sigma_{0}$ which intersects the cosmological and event horizons
(where $R=R_{1}$ and $R=R_{2}$, respectively). The initial data need
not correspond to the Kruskal-like coordinates $U,V$, but could be
in any other coordinates $\tilde{U},\tilde{V}$ which are regular
functions of respectively $U$ and $V$--- namely, we only require the
flux-free initial data to be regular through $\Sigma_{0}$. The
evolving solution then develops a vertex singularity at $P \equiv
(\tilde{U}_{0},\tilde{V}_{0})$, and a null IH singularity at
$\tilde{V}=\tilde{V}_{0}$. Now we add small (but finite), regular,
perturbations to the initial functions on $\Sigma_{0}$ such that
($\Psi,\Phi \neq 0$). Which properties of the flux-free singularity
will survive the perturbations and which will be modified?

We conjecture that the basic structure of a vertex singularity and a
null IH singularity is rather robust and will survive the
perturbation (see Fig. 3).
\begin{figure}[htb] \vspace{0.5cm} \epsfxsize=0.4\textwidth
\centerline{\epsffile{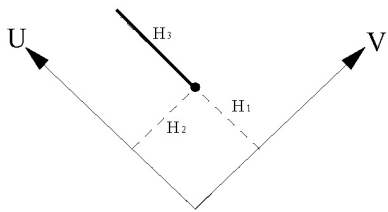}} \vspace*{0.3cm} \caption{\small
\bf \sl The general singularity structure. The three horizons are
displayed, denoted by $H_{i}$ ($i=1,2,3$). The singular IH is
displayed by a thick line emerging from the vertex singularity
(the intersection point of the three horizons).} \label{fig14}
\end{figure}
However, some of the features of the IH singularity will be modified.
In particular, the character of the flux-free IH singularity of being fully
locally-removable (by means of a local gauge transformation) will no longer hold
when fluxes are added. A more detailed description of the robust and non-robust
properties of the flux-free singularity is given below.

Our conjecture is based on several sources of evidence: (i) A linear
perturbation analysis \cite{unpub}. Note that at the linear level
the two fluxes $\Psi$ and $\Phi$ do not interact. The effect of each
flux may be separately explored by means of the exact Vaidya-like
solutions (see Appendix A), and then these two perturbations are
simply superposed. This drastically simplifies the analysis, as no
PDE needs be solved. In addition, the perturbation scheme suggests
that the effects of non-linear coupling between the two fluxes are
negligible compared to the basic effect of the blue-shifted incoming
flux. (This insight---namely the dominance of the linear
perturbation at the IH over the non-linear effects---has emerged in
previous studies of the IH singularity inside perturbed spherical
charged BHs \cite{Ori1, Burko2}, and, more generally, of
four-dimensional general-relativistic null spacetime singularities
\cite{Ori92,planar,flanagan}; and it applies to our toy semi-linear
system as well.) (ii) The general exact solution of our semi-linear
system for the case $h=cos(R)$ \cite{unpub}.

This case is exactly solvable, and it demonstrates all the elements
of the conjecture---except that it satisfies $k_{3}=k_{1}$ hence it
does not test issues related to the blue-shift phenomenon.
Nevertheless it provides support to our conjecture concerning all
properties of the vertex singularity, as well as some of the
properties of the IH singularity---in particular its null character.
(iii) Mathematical study of the general, flux-carrying solutions in
the case of a saw-tooth function $h(R)$ \cite{second}. In this case
$h(R)$ is made of three linear sections, patched together to form a
continuous but non-smooth function. This type of $h(R)$ confirms all
elements of the conjecture below--- for both cases $k_{3}=k_{1}$ and
$k_{3} \neq k_{1}$.

In what follows we present in detail our conjecture about the singularity structure
when the flux-free solution undergoes a generic small perturbation.
We divide the discussion into the vertex singularity and the IH singularity.

\subsection{The perturbed vertex singularity}

The perturbed solution exhibits a finite-time blow-up and
develops a vertex singularity at a slightly-modified point
$P' \equiv (\tilde{U}'_{0},\tilde{V}'_{0})$,
\footnote{Throughout this section we use a prime to denote perturbed quantities
and {\it not} the derivatives of functions}.
where
\begin{itemize}
  \item  $s$ diverges logarithmically to $ + \infty$ on approaching $P'$
  (e.g. from the past);
  \item The function $R$ is many-valued.
\end{itemize}
We stress again that this divergence of $s$ does not represent a
spacetime singularity --- it merely reflects the fact that the
proper time interval between $P'$ and any point to its past is
infinite. Therefore $P'$ functions as timelike infinity in the BH
spacetime. The presence of the singular point $P'$ {\it defines} the
horizons, and in this sense the very formation of the black hole.
More specifically, the cosmological and event horizons are now
defied to be the two null lines which intersect $P'$ at their future
[namely, $(\tilde{U}<\tilde{U}'_{0},\tilde{V}=\tilde{V}'_{0})$ and
$(\tilde{U}=\tilde{U}'_{0},\tilde{V}<\tilde{V}'_{0})$,
respectively]. These two null lines are no longer lines of constant
$R$. Nevertheless we conjecture that
\begin{itemize}
  \item  along each of these lines $R$ admits a well-defined value $R'_{i}$ ($i=1...2$)
  on approaching $P'$;
  \item  Furthermore these two values are both roots of the same function
  $H'(R)=h_{0}(R)+c'$,
with $c'$ close to the original $c$ (hence $R'_{1,2}$ are
close to the two original roots $R_{1,2}$).
\end{itemize}
\footnote{In fact, the same property is shared by $R'_{3}$, namely the limiting
value of $R$ on approaching $P'$ along the perturbed IH,
as discussed in the next subsection.}

{}From the GR point of view, this conjectured structure of the
perturbed vertex singularity is a manifestation of the {\it no-hair}
principle to our toy system; Namely all initial perturbations (the
non-vanishing fluxes) should disperse away and leave a black-hole
spacetime which at late time is well described by the corresponding
static solution.

\subsection{The perturbed IH singularity}

First we conjecture that, when a flux-free solution with an inner
horizon is weakly-perturbaed, no spacelike singularity will form in
the neighborhood of $P'$. Namely, in the neighborhood of $P'$ the
perturbed solution will be regular throughout
$\tilde{V}<\tilde{V}'_{0},\tilde{U}>\tilde{U}'_{0}$. However the
solution will generically develop a null singularity at the IH,
namely at $\tilde{V}=\tilde{V}'_{0}$ (for
$\tilde{U}>\tilde{U}'_{0}$). This singularity will be similar in
many respects to its flux-free counterpart:
\begin{itemize}
  \item  $R$ has a well-defined finite limit $R_{IH}(\tilde{U})$ at
  $\tilde{V} \to \tilde{V}'_{0}$;
   \item  Furthermore as $\tilde{U} \to \tilde{U}'_{0}$ (the $P'$ limit)
   $R_{IH}(\tilde{U})$ approaches a limiting value $R'_{3}$ which too is a root of
   $H'(R)$, close to the original root $R_{3}$.
  \item  Just as in the unperturbed case, the function $s$ generically {\it diverges}
  on the IH. The sign of divergence is determined by the relation
  between $k'_{1}$ and $k'_{3}$, where $k'_{i}$ is $|dH'/dR|$ at
  the root $R=R'_{i}$:
 Namely $s \to -\infty$ for $k'_{3}>k'_{1}$ and $s \to +\infty$ for $k'_{3}<k'_{1}$.
  \footnote{If the original flux-free solution satisfies
  $k_{3} < k_{1}$ or $k_{3} > k_{1}$,
  then we may assume that the perturbation is sufficiently small---and correspondingly
  $c'$ is sufficiently close to $c$---such that the same inequality is satisfied by the
  perturbaed quantities $k'_{1,3}$.
  Obviously in such a case, in the above condition for the sign of the divergent $s$ we
  may replace $k'_{1,3}$ by $k_{1,3}$.}
These two cases correspond to infinite blue shift or infinite red
shift, respectively.
  \item  In the special case $k'_{1}=k'_{3}$ we expect $s$ to remain regular at
  $\tilde{V} \to \tilde{V}'_{0}$. In this case the IH becomes a perfectly
  regular Cauchy horizon of our nonlinear hyperbolic system.
  \item  Returning to the generic case $k'_{1} \neq k'_{3}$: Although $s$ diverges,
  this divergence is locally removable by a gauge transformation
  \[\tilde{V}_{3}\equiv  =-(\tilde{V}'_{0}-\tilde{V})^{k'_{3}/k'_{1}} .  \]
  But this transformation leads to a divergence of $s$ along the cosmological
  horizon. Namely the divergence of $s$ at the IH is globally non-removable.
  \item  The indicator $\partial s / \partial \tilde{U}$ is regular at the IH limit
  $\tilde{V} \to \tilde{V}'_{0}$. As discussed in the previous section,
  this indicates that indeed the divergence of $s$ is locally removable.
\end{itemize}

All the above features merely reflected the similarity between the
perturbed and unperturbed (flux-free) IH singularities. But there
also is an important difference between the perturbed and
unperturbed singularities in the case of infinite blue shift. This
difference is manifested by the following properties which all hold
for generically-perturbaed solutions with $k'_{3} >2 k'_{1}$
~\footnote{This condition was derived in the perturbed RNDS case in
Ref. \cite{Brady3}. }(to which we may refer as ``strongly-divergent
blue shift''):
\begin{itemize}
  \item The quantity
  $e^{-s} (\partial R / \partial \tilde{V})$ diverges at the IH limit
  $\tilde{V} \to \tilde{V}'_{0}$.
  \item   The same applies to the scalar quantity
   $e^{-s} (\partial R / \partial \tilde{V}) (\partial R / \partial \tilde{U})$.
  \item  As a direct consequence
  of the above
  the mass function,
  which was fixed (M=-c) in the flux-free case,
  diverges at the IH of the generically-perturbed solution.
  \end{itemize}
The meaning of these last indicators is simple: The divergence of
$s$ at the IH may be locally removed by a gauge transformation as
mentioned above. This transformation renders the solution continuous
at the IH. Yet generically the flux-carrying solution fails to be
smooth at the IH. This lack of smoothness is manifested by the
divergence of certain gauge-invariant quantities, so it cannot be
cured by any gauge transformation. In the case of strongly-divergent
blue shift $(k'_{3} >2 k'_{1})$ the solution fails to be $C^{1}$. As
it turns out, in the case
 $ k'_{1}<k'_{3} <2 k'_{1}$ the
solution is $C^{1}$ but fails to be $C^{2}$ at the IH. In the case
$k'_{3} < k'_{1}$ (infinite red-shift) the solution will be $C^{2}$,
yet generically smoothness will fail at a certain derivative orders
(unless $2 k'_{1} / k'_{3}$ is an integer). \footnote{The general
rule is the following: The conjectured smoothness level of $R$ as a
function of $\tilde{V}_{3}$ is the same as that of the function
$|\tilde{V}_{3}|^{2k'_{1}/k'_{3}}$.}

In both blue-shift cases ($k'_{3} >2 k'_{1}$ and $ k'_{1}<k'_{3} <2
k'_{1}$) the generally-perturbed solution fails to be $C^{2}$ at the
IH. In General-Relativistic terms this means that the Reimann tensor
will diverge there (e.g. as measured by polynomial scalars or by its
p.p. components)---even if the mass function remains bounded. On the
other hand, in the red-shift case $(k'_{3} <2 k'_{1})$ the Riemann
tensor will remain bounded, though generically its derivatives will
blow up at a certain order.

Our primary concern here is the case of generic perturbations, in
which both ingoing and outgoing fluxes are present. However, it is
also interesting to consider the case a BH perturbed by an ingoing
flux while the outgoing flux vanishes. This case is exactly solvable
by means of the Vaidya-like solution (see Appendix). In this case
scalar quantities as $M$ and $e^{-s} (\partial R / \partial
\tilde{V}) (\partial R / \partial \tilde{U})$ remain bounded.
However $e^{-s} (\partial R / \partial \tilde{V})$ generically
diverges at the IH if $k'_{3} >2 k'_{1}$. Although this quantity is
not a scalar, its divergence is a gauge-invariant phenomenon as
discussed in the previous section. Again this divergence indicates
an unbounded curvature (though only in the p.p. sense).

As was mentioned above, the divergence of $s$ at the IH is
conjectured to be removable by means of a local gauge
transformation, leading to a continuous (though not smooth) solution
at the IH. From the GR point of view this means that the metric
tensor has a continuous (though not smooth) non-singular limit at
the IH. This means that the IH singularity is weak (namely
physically non-destructive), as discussed at the Introduction.

We summarize some of the conjectured properties of the
generically-perturbed singularity in the table below.

\begin{table}[h]
\vspace*{0.3cm}
\begin{tabular}{|c||c|c|c|c|} \hline
 {\emph{The quantity}} & \multicolumn{3}{c | }{\emph{Type of divergence}} & \emph{Condition for}       \\
  &\multicolumn{3}{c | }{} & \emph{the divergence}
 \\ \hline \hline
  & no null fluid & only influx & two null fluids & \\
  \cline{2-4}
     & $+\infty$ & $+\infty$ & $+\infty$ & $k_{1}>k_{3}$ \\
 $s$ &  --- &--- & --- & $k_{1}=k_{3}$\\
     & $-\infty$ & $-\infty$ & $-\infty$ & $k_{3}>k_{1}$ \\ \hline
$e^{-s}R_{,v}$ & --- & $+\infty$ & $+\infty$ & $k_{3}>2k_{1}$ \\
\hline $e^{-s}R_{,v}R_{,u}$ & --- & --- & $+\infty$ & $k_{3}>2k_{1}$
\\ \hline
\end{tabular}
\caption{\small \bf \sl The divergence of various quantities at the
inner horizon $H_{3}$ expected to be found in various cases.}
\end{table}

\section{Future research directions}

The problem which motivated this paper concerned the internal
structure of generically-perturbed spinning vacuum black holes. One
would like to prove (or disprove) the generic formation of a null
singularity inside the BH, and its weakness. We find it likely that
our simple two-dimensional semi-linear system presented here
properly mimics these properties of General-Relativistic spinning
black holes.

{}From this perspective the task of mathematically exploring the
internal structure of realistic spinning black holes may be divided
into the following stages: (i) Proving (or disproving) the
conjecture presented in the previous section concerning the generic
behavior of our semi-linear system; (ii) Extending the analysis to
the asymptotically-flat case, namely to functions $H(R)$ with two
roots and with large-$R$ asymptotic behavior similar to the
$\Lambda=0$ case of Eq. (\ref {HRNDS}); and (iii) Further extending
the analysis to the real problem of a generically-perturbed spinning
black hole.

Although stages (i) and (ii) are not easy, stage (iii) will probably pose a much harder challenge,
as it involves the transition from two to four dimensions (and with a much larger number of
unknown functions). Nevertheless it may be hoped that the insight gained from our simple
toy system in stages (i,ii) may make this challenge a bit easier.

\

\noindent{\bf Acknowledgements:} We thank G. Wolansky for many
discussions and comments on the manuscript. In addition we thank J.
Feinberg for comments on a preliminary version of the manuscript.

\appendix
\section{The case of one flux (Vaidya-like solution)}
\label{vaidya} In the case of a flux in one direction only our
semi-linear system may be reduced to a single ordinary differential
equation along null lines. Consider for example the case $\Phi=0$,
namely outflux only (the other case $\Psi=0$ may be treated in a
fully analogous manner). Then Eq. (\ref{constraint v}) yields
$R_{,v}=c_{u}(u) e^{s}$. After a re-labeling of $u$ this becomes \be
e^{s}=R_{,v} \ , \label{s invaidya}\ee where we have considered here
the case of positive $c_{u}$ for concreteness (the case $c_{u}<0$
proceeds in a similar manner). Substituting this in the right-hand
side of the first evolution equation in (\ref{evh}) we obtain
\[R_{,uv}=-R_{,v}h'(R).\] Integration with respect to $v$ gives
\be R_{,u}=-h_{0}(R)+M_{u}(u)\label{vai}\ee where $M_{u}(u)$ is an
arbitrary function of $u$.

Differentiation of Eq. (\ref{s invaidya}) with respect to $u$,
substitution of the evolution equation for $R$ and then a second
differentiation with respect to $v$, one finds that the second
evolution equation (\ref{evh}) is satisfied as well.

Thus, the original system (\ref{evh}) of evolution equations has been reduced
to a single ordinary differential equation ($\ref{vai}$) along lines of constant $v$.
Solving this ODE (with a $v$-dependent initial condition $R_{0}(v)$
at a certain initial $u$ value
$u=u_{0}$) yields the function $R(u,v)$. Subsequently $s$ is obtained by
Eq. (\ref{s invaidya}).
\footnote{Without loss of generality one may take $R_{0}(v)=\pm v$ or
$R_{0}(v)=const$ (at least piecewise).
Any other (monotonic, non-constant) function $R_{0}(v)$
may be brought, via a gauge transformation of $v$,
into $R_{0}=v$ or $R_{0}=-v$, depending on whether the
original function $R_{0}(v)$ is increasing or decreasing.}

Substituting this solution in Eq. (\ref{Mdef}) one obtains
\[M(u,v)=M_{u}(u) \ . \]
The outflux is then given by Eq. (\ref{constraint u}) or
(\ref{Mdw}), \be\Psi(u)=\frac{dM_{u}}{du} \ee (and, recall,
$\Phi=0$).

\end{document}